\documentclass[twocolumn,showpacs,preprintnumbers,amsmath,amssymb]{revtex4}

\usepackage{graphicx}
\usepackage{dcolumn}
\usepackage{bm}

\begin{document}

\preprint{APS/Fluct-QChains}

\title{Work fluctuations in quantum spin chains}

\author{Sven Dorosz}
 \affiliation{Department of Physics, Virginia Polytechnic Institute and State University, Blacksburg, Viriginia 24061-0435, USA}
 
 \author{Thierry Platini}%
\affiliation{Laboratoire de Physique des Mat\'eriaux, 
	UMR CNRS 7556,\\ 
  Universit\'e Henri Poincar\'e, Nancy
  1,\\ B.P. 239,
  F-54506  Vand\oe uvre les Nancy Cedex, France
}%

\author{Dragi Karevski}
 \email{karevski@lpm.u-nancy.fr}
\affiliation{Laboratoire de Physique des Mat\'eriaux, 
	UMR CNRS 7556,\\ 
  Universit\'e Henri Poincar\'e, Nancy
  1,\\ B.P. 239,
  F-54506  Vand\oe uvre les Nancy Cedex, France
}%

\date{\today}

\begin{abstract}
We study the work fluctuations of two types of finite quantum spin chains under the application of a time-dependent magnetic field in the context of the fluctuation relation and Jarzynski equality. The two types of quantum chains correspond to the integrable Ising quantum chain and the non-integrable XX quantum chain in a longitudinal magnetic field. For several magnetic field protocols, the quantum Crooks and Jarzynski relations are numerically tested and fulfilled. As a more interesting situation, we consider the forcing regime where a periodic magnetic field is applied.  In the Ising case we give an exact solution in terms of double confluent Heun functions. We show that the fluctuations of the work performed by the external periodic drift are maximum at a frequency proportional to the amplitude of the field. In the non-integrable case, we show that depending on the field frequency a sharp transition is observed between a Poisson limit work distribution at high frequencies toward a normal work distribution at low frequencies.
\end{abstract}

\pacs{Valid PACS appear here}

\maketitle

\section{Introduction}
The study of fluctuations in non-equilibrium small systems has become an active field of research during the last years. The reasons are twofold: on one hand, nanoscaled systems are nowadays quite easily manufactured, opening the door to the emergence of various nano-technologies. On the other hand in the field of non-equilibrium statistical mechanics, where exact results are very few, the discovery of fluctuation symmetries \cite{evan1,evan2} expressed by the Gallavotti-Cohen fluctuation theorem \cite{gaco,kurc,lesp,maes,seif,hasa} and Jarzynski equality \cite{jarz1} have opened new theoretical perspectives. The fluctuation theorem is a statement on the time-reversal symmetry of the fluctuations of the entropy production along a non-equilibrium path \cite{maes}. Whereas this theorem is an asymptotic statement, Crooks derived an interesting identity reading \cite{croo1,croo2}  
\begin{equation}
\frac{P_F(\Delta S)}{P_B(-\Delta S)}=e^{\Delta S}\; .
\end{equation}
$\Delta S$ is the entropy production for a system driven during a time $\tau$ within a forward protocol $\lambda_F(t)$,  $P_{F}$ is the distribution of entropy production in the forward process and $P_B$ is the distribution associated to the backward process, that is when the system is driven in a time-reversed manner. The celebrated Jarzynski equality \cite{jarz1} is easily derived from the Crooks relation. Indeed, utilizing for a thermalised system $\Delta S=\beta(W-\Delta F)$, with $\Delta F=[F(\lambda_F(\tau))-F(\lambda(0))]$ and integrating over $W$ we have 
\begin{equation}
\langle e^{-\beta W}\rangle = e^{-\beta \Delta F}\; 
\label{eqjarz1}
\end{equation}
which is Jarzynski relation.
These relations, initially derived for classical systems, have been  extended to the quantal world as well \cite{kurc2,mota,monn,muka,esmu,dema} within various setup and prescriptions for the actual measurement of work performed on a quantum system \cite{alni}. 

As it is clear from the extensivity of the entropy production, the probability to observe a decrease of entropy in a given non-equilibrium trajectory is exponentially small for a macroscopic system. However, for microscopic or mesoscopic systems these untypical trajectories arise with a significant probability and consequently can be observed and measured in actual experimental setup. One may mention in particular experimental tests on the stretching of RNA molecules \cite{lidu,cori},  experiments on torsional pendulum \cite{doci}, on colloidal particles \cite{wase,care}, on photochromic defect center in diamonds \cite{tisc}. See Ref. \cite{rito} for an excellent review of the experimental related investigations. It is then interesting to calculate or predict the shape of these fluctuations in some explicit models for small system size \cite{kare1,chka}. In this contribution we focus our attention to the work fluctuations of two time-dependent quantum spin chains where the time dependence is due to an externally applied magnetic field. The two models considered are the integrable quantum Ising chain in a time-dependent transverse field and the $XX$-quantum chain with a longitudinal magnetic field that breaks its integrability.
We consider these two cases as archetypical of the integrable and non-integrable situations and we expect that the work distributions will reflect somehow these differences. We check Jarzynski and Crooks relations and compute explicitly the non-equilibrium work distributions associated to several magnetic field protocols.  In particular, we study the limiting steady distributions in the important case of periodically driven system. 

The paper is organized as follows: in the next section we define the models and protocols for studying work fluctuations.  We insist in particular on the definition of work we are using in our study, pointing explicitly as it was done in \cite{alni,talu} the
discrepancies between this definition and the use of a work operator. We present then the numerical results corroborating Jarzynski and Crooks relations before turning to the main part which is the study of work fluctuations within a periodic drift. An exact solution, in terms of double confluent Heun functions is given in the Ising case. The study of the $XX$-chain is only numerical, but nevertheless evidences are drown out to show clearly the appearance of a sharp transition between an exponential work distribution at high field frequencies toward a gaussian work distribution at lower frequencies.
We summarize and discuss our results in the last section.

\section{Quantum spin chains in a time dependent field}
\subsection{Definition of the model and protocol}
We study quantum chains with a time-dependent field, in the context of fluctuations of the work performed on them by the time-dependent force, defined through the following Hamiltonian:
\begin{equation}
H=-\frac{1}{2}\sum_{i=1}^{N-1}\sigma^x_i\sigma^x_{i+1}
-\frac{\Delta}{2}\sum_{i=1}^{N-1}\sigma^z_i\sigma^z_{i+1}
-\frac{1}{2}h(t)\sum_{i=1}^{N}\sigma^z_i
\label{hamil1}
\end{equation}
in particular in the two cases $\Delta=0$ and $\Delta=1$. 
The $\sigma$'s are Pauli's matrices and $h(t)$ is a time-dependend magnetic field applied in the $z$-direction and leading to the Zeeman term $-h(t)M^z$ where $M^z$ is the total $z$-component of the magnetization. 

At $\Delta=0$, it corresponds to 
the Ising quantum chain. In a static transverse field the Ising chain can be mapped after a Jordan-Wigner transformation onto a fermionic problem which can be diagonalised after a suitable canonical (Bogoliubov) transformation.  In the thermodynamic limit $N\rightarrow\infty$, this model presents a quantum phase transition at $h=h_c=1$ which is in the 2d classical Ising model universality class. The integrability and non-triviality (in the sense of physical properties) of the model is at the origin of its wide  use in many fundamental studies concerning in particular the testing grounds of non-equilibrium quantum statistical mechanics. One may mention for example relaxation properties from an inhomogeneous initial state \cite{sctr,kare2,kare3,aschba1,aschba2,ogat}, entropy production after a local quench \cite{caca,kare4}, entropy production in the unique NESS (non-equilibrium steady state) generated from a two temperature initial state \cite{aschba3}, or even relaxation properties after a quench through the critical point (basically $h$ is varied from $h_i>h_c$ to $h_f<h_c$) \cite{seng,cherng}.  

The case $\Delta=1$ corresponds to the isotropic XX-quantum chain in a time-dependent magnetic field, which breaks the integrability of the model. 
With a periodic dependence of the field,
this model has to be compared to the quantum kicked Heisenberg chain:
\begin{equation}
H(t)=-\frac{1}{2}\sum_{j=0}^{L-1}\left(\sigma^x_i\sigma^x_{i+1}
+\sigma^z_i\sigma^z_{i+1}+\delta(t)V \sigma^y_i\sigma^y_{i+1}\right)
\end{equation}
where the time-dependent perturbation is periodically switched on. This model was studied in the context of the transition from integrability to ergodicity in the thermodynamic limit \cite{prosen1,prosen2,mont}. Here instead of the two-body interaction (in the fermionic picture) $\sigma^y_i\sigma^y_{i+1}$ we apply a magnetic field in the $z$-direction which leads to a many-body fermionic term.

In the following, we consider the distribution of the work performed by the time-dependent field $h(t)$ from the initial time $t_i$ to the final time $t_f$. The varying field will lead to transitions between the initial state of the system to a new state at the final time. Here we use as a definition for the work the difference of energies in the final state and initial state, $W_{if}=\Delta E_{if}=E_f-E_i$, which assumes that we have measurements performed at $t_i$ and $t_f$.
The work distribution $P(W)$ is a weighted sum over all initial and final states of the quantum transition probability $|(\phi_f,U\phi_i)|^2$, where $U$ is the unitary time-evolution operator associated to the time-dependent Hamiltonian $H(t)$, and is given by
\begin{equation}
P(W)=\sum_{i,f}\delta(\Delta E_{if}-W) |(\phi_f,U\phi_i)|^2\; \omega^{\phi_i}
\label{eqwork}
\end{equation}
where $\phi_i$ and $\phi_f$ are respectively initial and final eigenstates of the initial and final Hamiltonian and $\omega^{\phi_i}=(\phi_i,\rho_0\phi_i)$ where $\rho_0\propto e^{-\beta H(0)}$ is the initial (which is supposed to be canonical here) density matrix. 
It is clear that this definition differs from the introduction of a work operator defined such that
its expectation in the state $\rho$ would give the actual performed work:
\begin{equation}
W=\int_{t_i}^{t_f} {\rm d}s \; Tr\{\rho(s)\frac{d H}{ds}\}
\end{equation}
which can be shown to be given by
\begin{eqnarray}
&W=Tr\{\rho(t_i) \delta H\}=\nonumber\\
&Tr\{\rho(t_i)[U^+(t_f,t_i)H(t_f)U(t_f,t_i)-H(t_i)]\}
\end{eqnarray}
after a short algebra.  One is then tempted to define as a work operator  $\delta H=\tilde{H}(t_f)-H(t_i)$, with $\tilde{A}$ the Heisenberg picture of $A$. 
However, it is clear that in general the moments  $\langle (\delta H)^n\rangle$ would differ from the moments $\langle W^n\rangle=\int {\rm d} W\; W^n P(W)$ with $P(W)$ defined in (\ref{eqwork}) and consequently the expectation of the operator 
$e^{-\beta \delta H}$ 
\begin{equation}
\langle e^{-\beta\delta H}\rangle = Tr \{ \rho(t_i) e^{-\beta \delta H}\}
\end{equation}
will not necessarily give the Jarzynski result $e^{-\beta \Delta F}$.
Nevertheless,  one can show that the moments so defined will differ only for $n>2$, so that for the average and average square work one can use the work operator $\delta H$ since then 
\begin{eqnarray}
\langle \delta H\rangle = Tr\{\rho(t_i) \delta H\}=\int {\rm d} W \; W P(W)\nonumber\\
\langle (\delta H)^2\rangle = Tr\{\rho(t_i) (\delta H)^2\}=\int {\rm d} W \; W^2 P(W)
\end{eqnarray}
with $P(W)=\sum_{i,f}\delta(\Delta E_{if}-W) |(\phi_f,U\phi_i)|^2\; \omega^{\phi_i}$.
The equality of the first moments is trivial. To prove the equality for the second moments, one may notices that for $[\rho(t_i),H(t_i)]=0$ using the cyclicality property of the trace one has $Tr\{\rho(t_i)[\tilde{H}(t_f),H(t_i)]\}=0$, which then leads trivially to the equality of the second moments: 
$\int dW W^2 P(W)=\sum_{i,f}(E_f-E_i)^2|(\phi_f,U\phi_i)|^2\; \omega^{\phi_i}=
Tr\{\rho(t_i)(\tilde{H}(t_f)-H(t_i))^2\}-Tr\{\rho(t_i)[\tilde{H}(t_f),H(t_i)]\}=
Tr\{\rho(t_i)(\tilde{H}(t_f)-H(t_i))^2\}
$.
In short, the second moments agree whatever the commutator $[\tilde{H}(t_f),H(t_i)]$ is. At the level of the third moments, $n=3$, one may show that a non-vanishing term $Tr\{\rho(t_i)[\tilde{H}(t_f),H(t_i)]\tilde{H}(t_f)\}$ comes into play leading to a difference between the moments associated to the two preceding definitions of the work.

In order to compute the work distribution one needs to know the unitary time-evolution operator $U$ for the given field protocol, which is a difficult task in general. The numerical study assumes that the true dynamics is well approximated by the step like unitary evolution 
\begin{equation}
U(t,t_0)=\prod_{n=0}^{M-1}U(t_{n+1},t_{n})
\label{eqU1}
\end{equation}
with 
\begin{equation}
U(t_{n},t_{n-1})=e^{iH(t_n)\Delta t}
\label{eqU2}
\end{equation}
where $\Delta t$ is an elementary time increment, meaning that the real function $h(t)$ is approximated by a step function. The true dynamics is obtained by taking the limit $\Delta t\rightarrow 0$, and $M\rightarrow \infty$. In the present study we  use typically values of $\Delta t$ of order $10^{-2}$ (time is measured in units where $\hbar=1$ and  the spin-spin coupling $J^x$ in the $x$-direction  of the spin chains is set to one).
The transition probabilities $w_{i\rightarrow f}=|(\phi_i, U(t,t_0),\phi_f)|^2$, where $\phi_{i}$ ($\phi_f$) are eigenstates of the initial(final) Hamiltonian, are computed numerically using (\ref{eqU1}) and (\ref{eqU2}) on small chains of typically less than $10$ spins and with open boundary conditions.

\subsection{Jarzynski and Crooks relations}
In order to test our numerical procedure we compare first the free energy differences obtained from the Jarzynski equality (\ref{eqjarz1})
 in a linear varying field protocol with the equilibrium free energy differences calculated directly from the partition function
\begin{equation}
F(h(t))=-\frac{1}{\beta}\ln Tr\{e^{-\beta H(t)}\}\; .
\end{equation}
\begin{figure} [ht]
\includegraphics[width=8cm]{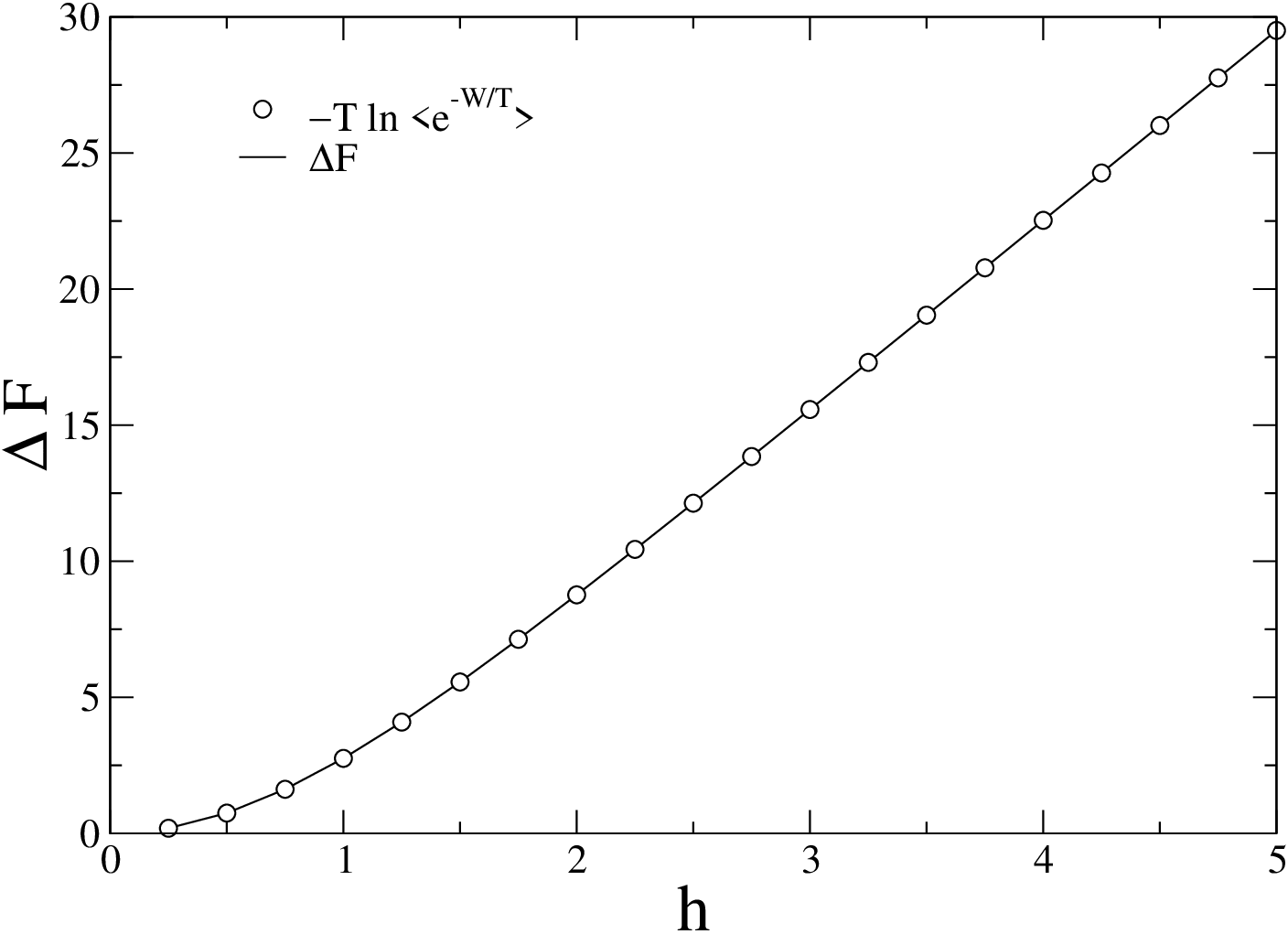}
        \caption{Free energy difference obtained from the Jarzynski equality compared to the exact value for the Ising chain with $N=7$ as a function of the terminal field value at $\beta=1$ and $h(t)=t$.  
        \label{Fig1}  }
\end{figure}

In figure 1 we have plotted the free energy differences obtained from the Jarzynski equality and from a direct equilibrium calculation as a function of the final field value $h$ at inverse temperature $\beta=1$. The agreement is excellent with a relative deviation which is less than $10^{-6}$. 

We have also considered the ratio $P_F(W)/P_B(-W)=P(W)/P(-W)$ for the symmetric protocol $h(t)=h_o \sin(\pi t/\tau)$ for $t\in[0,\tau]$. In figure 2 we see explicitly the exponential dependence  $e^{\beta W}$ for $h_o=1/2$ and $2$ for a chain of size $N=8$. Again, the agreement of the numerics with the analytical Crooks and Jarzynski relations is fulfilled with a relative deviation which is less than $10^{-6}$. 
We have tested also the Crooks relation in the non-symmetrical protocol whose results are presented in  figure 3. Again, the exponential dependence $e^{\beta W}$ is clearly seen on the graph. 
\begin{figure} [ht]
        \includegraphics[width=8cm]{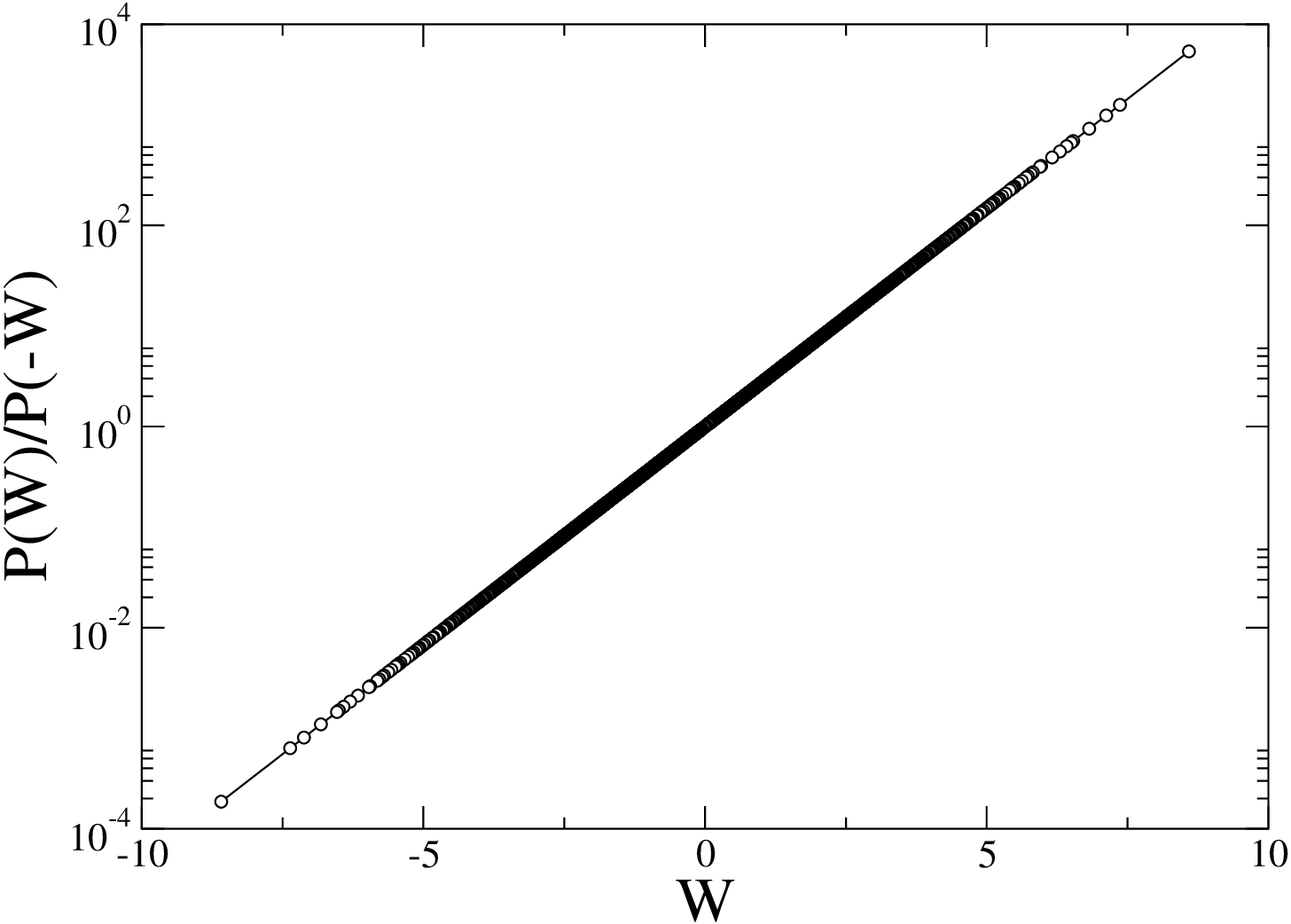}
        \caption{Crooks fluctuation ratio in the symmetrical case for the XX chain at $\Delta=1$ with $N=8$ at $\beta=1$ and $\tau=1$.  
        \label{Fig2}  }
        \end{figure}

\begin{figure} [ht]
     \includegraphics[width=8cm]{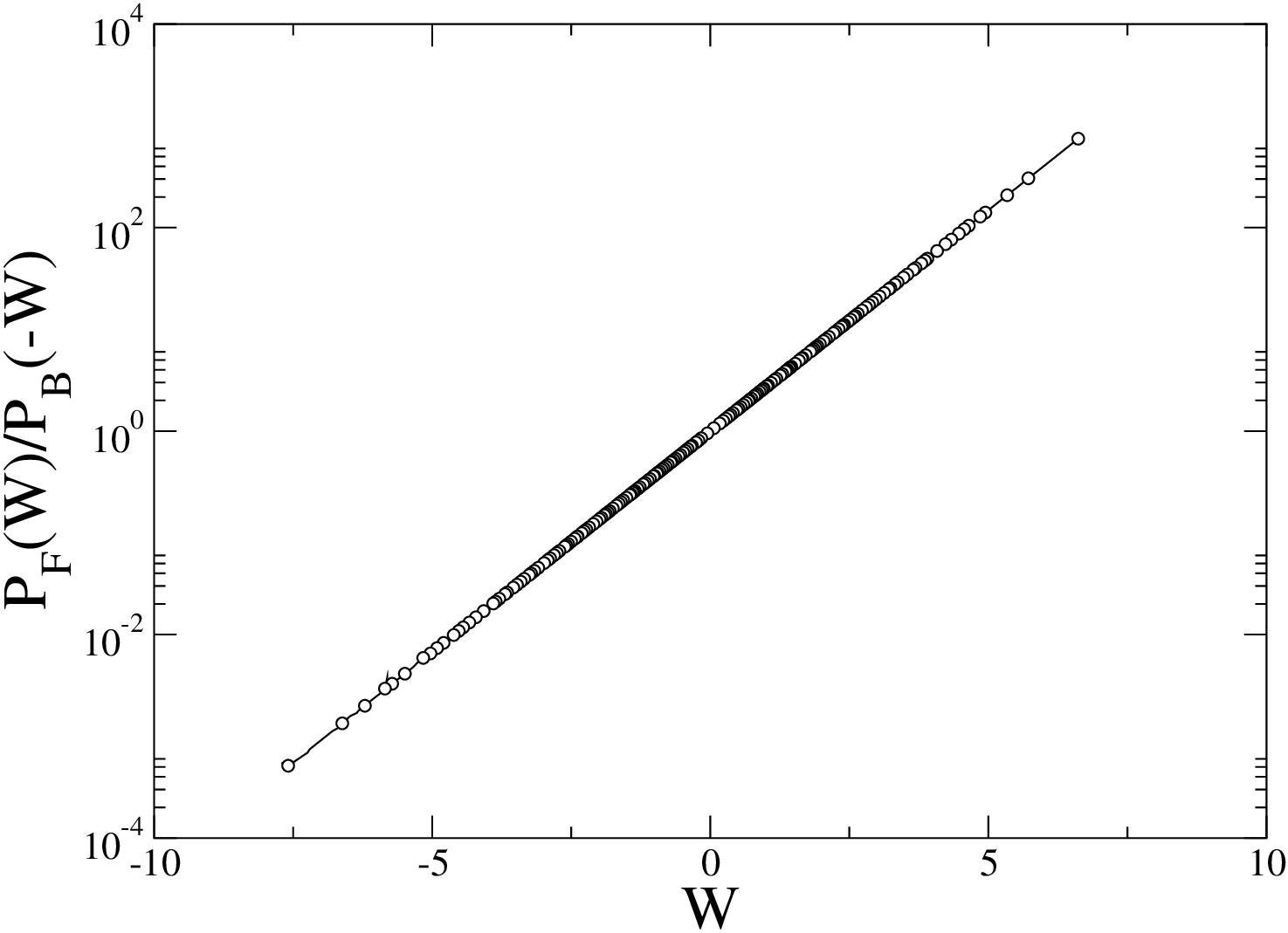}
      \caption{Crooks fluctuation ratio in the asymmetrical case for the XX chain at $\Delta=1$ with $N=8$ at $\beta=1$.  
        \label{Fig3}  }
\end{figure}

\subsection{Stationary distributions in the driven regime}
\subsubsection{Ising quantum chain}
\paragraph{Numerical study}
We consider now the case where the system is forced with  a periodic external field with period $\tau$, maximum value $h_o$ and minimum value $0$.
We are interested in the limiting stationary work distribution (if any) obtained after many periods of the external driving field.

\begin{figure} [ht]
       \includegraphics[width=8cm]{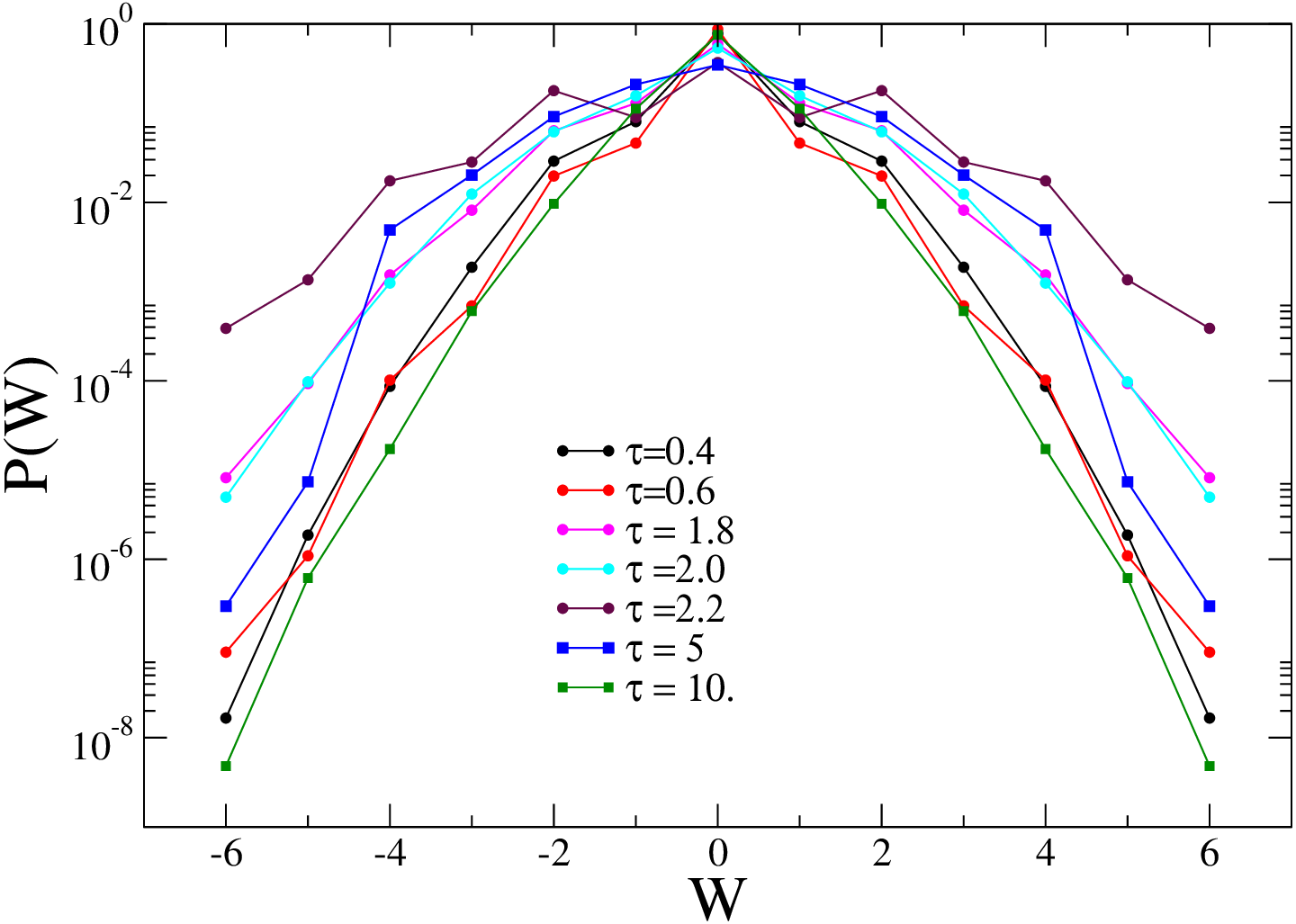}
        \caption{Work probability distribution for the Ising model at with $N=7$ spins at $\beta=0$ for $h_o=1/2$.  
        \label{Fig4}  }
\end{figure}

\begin{figure} [ht]
     \includegraphics[width=8cm]{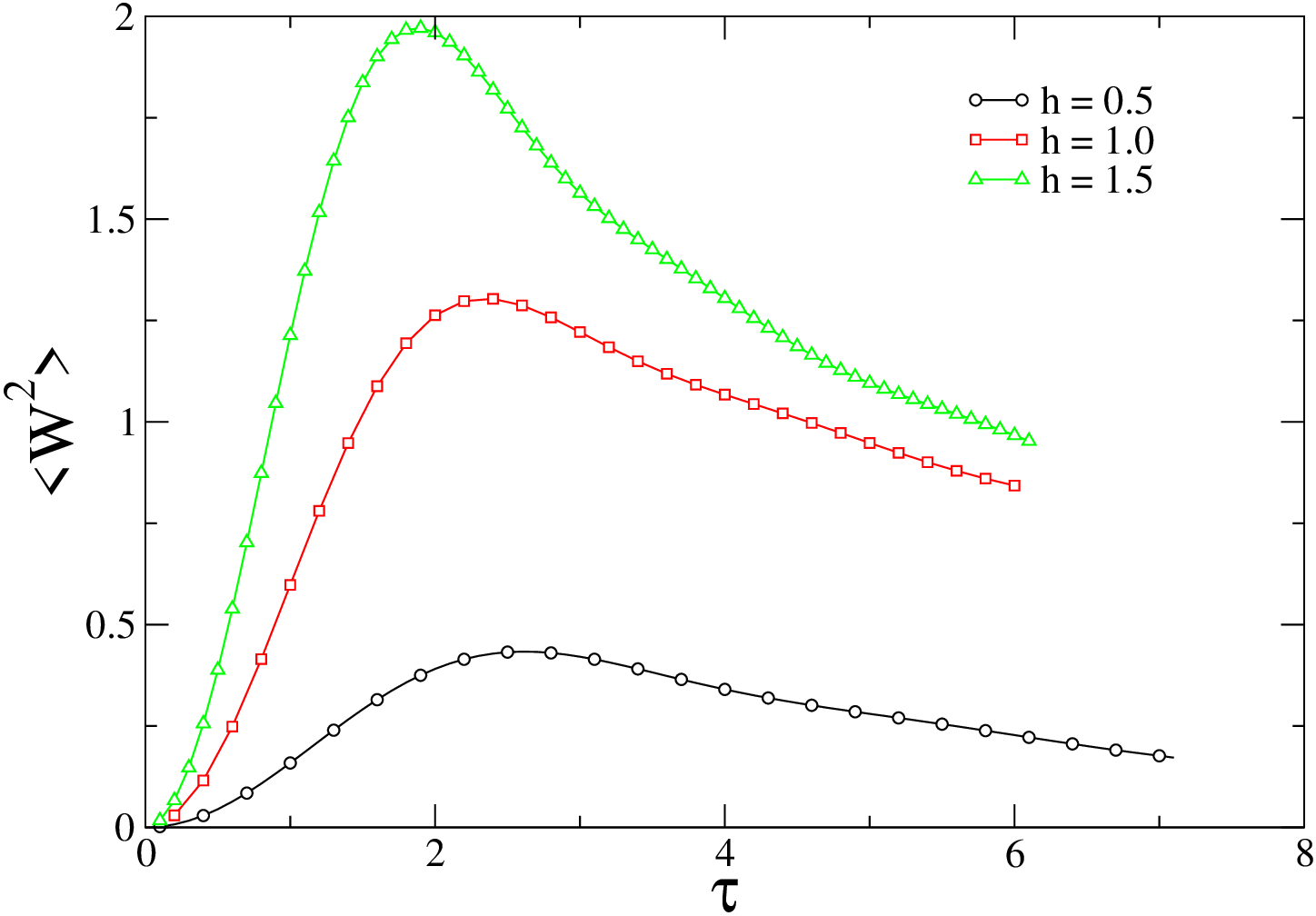}
              \caption{$\langle W^2\rangle$ for the Ising model with $N=7$ spins  at $\beta=0$ as a function of the period $\tau$. 
        \label{Fig5}  }
        \end{figure}

Before going on the driven situation, we concentrate on the fluctuations of the work after one period.
In figure \ref{Fig4} we show the distributions obtained on the Ising chain with $N=7$ spins for $h_o=1/2$ and for different periods $\tau$ of the forcing field for an initial infinite temperature state $\beta=0$ (equidistribution of the initial eigenstates). 
In that case, obviously, the average work is vanishing, which is reflected in the symmetry of the work distribution $P(W)=P(-W)$. On figure \ref{Fig4}
we see clearly that the evolution of the distribution is not monotonic with the forcing period $\tau$. The width of the distribution first increases with $\tau$, reaching a maximum and then decreases with $\tau$. 
This behavior is clearly seen on the second moment $\langle W^2\rangle$ in figure \ref{Fig5} where we show the variance $\langle W^2\rangle_{\beta=0}$ as a function of $\tau$ for different maximum field values $h_o$ for a $N=7$ spin chain. We see that the width of the distribution increases with the amplitude of the perturbation, since then the time-dependent perturbation is more effectively coupled with the spin chain. Moreover, the period $\tau_{max}$ associated to the maximum width decreases as the amplitude is increased with, as seen from the numerics, a law $\tau_{max}\sim 1/h$ for large fields enough ($h_o>1$). Moreover we have seen numerically that $\tau_{max}$ is almost independent of the system size. As the field is increased, the maximum value of the variance reaches an asymptotic finite value which is linearly depending on the size $N$ of the chain. One has from the numerics $\langle W^2\rangle_{\beta=0,max}=(N-1)/2$.

\begin{figure} [ht]
        \includegraphics[width=8cm]{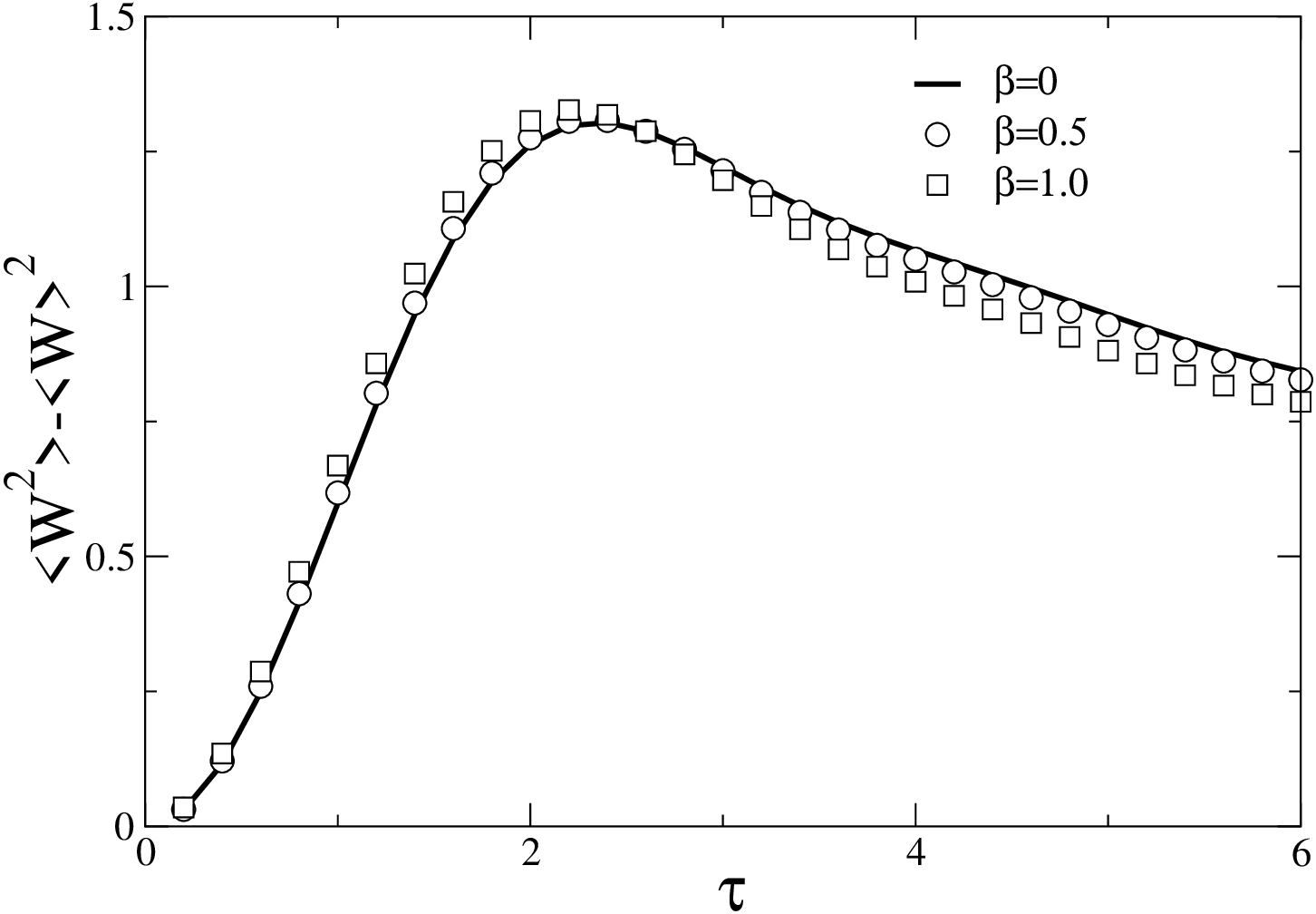}
       \caption{Variance of the work distribution for the Ising model with $N=7$ spins at $h_o=1$ as a function of the period $\tau$ for $\beta=0$, $\beta=1/2$ and $\beta=1$. 
        \label{Fig6} }
\end{figure}

We have also computed the first and second moments of the distribution as a function of $\tau$ for different inverse temperatures $\beta$. The initial state temperature acts as a scale factor for the moments:
\begin{equation}
\langle W\rangle_\beta= \tanh(\beta/2) \langle \delta W\rangle
\label{Wtanh1}
\end{equation}
where $\langle \delta W\rangle$, defined by this equation,  is the temperature independent part of the average work. As seen on figure \ref{Fig6} one has for the variance
\begin{equation}
\langle W^2\rangle_\beta-\langle W\rangle^2_\beta \simeq\langle W^2\rangle_0\; 
\end{equation}
meaning that the width of the distribution has a very small dependence on the initial state temperature, at least in the considered range of $\beta$. 

\begin{figure} [ht]
        \includegraphics[width=8cm]{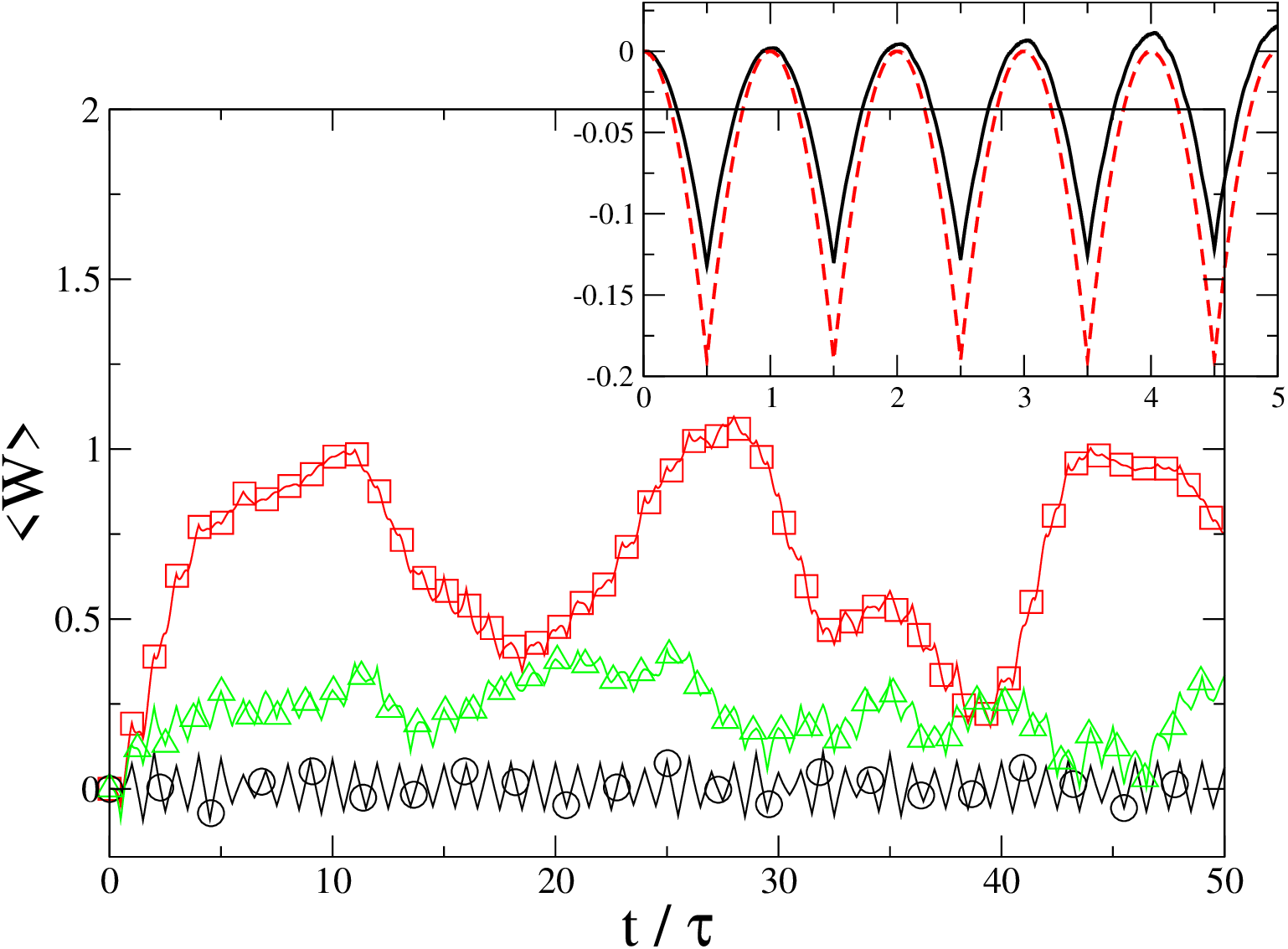}
       \caption{Time evolution of the average work for the Ising chain with a triangular periodic transverse field at $\beta=1$ and maximal amplitude $h_o=1/2$ with periods $\tau=1$ (circles), $\tau=3$ (squares) and $\tau=5$ (triangles). The inset shows the free energy difference (lower curve) and the average work (top curve) for a large period $\tau=50$ of the field. In this case the transformation is nearly adiabatic.    
        \label{Fig7} }
\end{figure}

We turn now to the driven situation with many periods. In figure \ref{Fig7} we present the time evolution  of the average work for different periods $\tau$ at a field value $h_o=1/2$ and inverse temperature $\beta=1$. As expected, for large periods the system follows almost adiabatically the variation of the associated equilibrium free energy (the reversible work). Moreover, it is seen as expected from the second principle that the actual average work is always greater than the free energy difference (see the inset). As the period of the external transverse field is lowered, we see that the deviation from the free energy difference becomes larger and one is no more able to recognize the underlying oscillatory external field.   
The largest variations and most "chaotic" behavior is obtained at the value $\tau_{max}$ previously identified for one period. At higher frequencies the fluctuations of the average work decrease again.

\paragraph{Exact solution for the driven Ising model}
With a sinusoidal forcing field $h(t)\propto \sin(\omega t)$ it is in fact possible to solve exactly the dynamics of the Ising quantum chain. Indeed, following the lines of Ref.\cite{bamc}, we can put the periodic boundary conditions Hamiltonian (\ref{hamil1}), after a Jordan Wigner transformation followed by a Bogoliubov one, into a sum of commuting operators:
\begin{equation}
H=\sum_{p=1}^{N/2} \tilde{H}_p
\end{equation} 
with
\begin{eqnarray}
\tilde{H}_p=(\cos\phi_p-h(t))[a^+_p a_p+a^+_{-p}a_{-p}]\\
-i\frac{1}{2}\sin\phi_p[a^+_p a^+_{-p}+a_p a_{-p}]+h(t)
\end{eqnarray}
where $[\tilde{H}_p,\tilde{H}_q]=0$ and where $\phi_p=2\pi p/N$. The $a^+,a$ are Fermi operators in momentum space. Using the following basis $(|0\rangle, a^+_p a^+_{-p}|0\rangle, a^+_p|0\rangle, a^+_{-p}|0\rangle)$, where $|0\rangle$ is the vacuum of the $a$ fermions, we have the $4\times 4$ representation 
\begin{equation}
\tilde{H}_p= \left(
\begin{array}{cccc}
h(t) & -i\sin\phi_p & 0 & 0\\
i\sin\phi_p&2\cos\phi_p-h(t)&0&0\\
0&0&\cos\phi_p&0\\
0&0&0&\cos\phi_p
\end{array}
\right)
\label{HpIsing}
\end{equation}
In the Heisenberg picture, the time evolution matrix $U_p(t)$ in the subset $p$ is governed by the equation ($\hbar=1$)
\begin{equation}
i \frac{d}{dt} U_p(t)=U_p(t)\tilde{H}_p(t)
\label{dif1}
\end{equation}
with the boundary condition $U_p(0)=I$. Using (\ref{HpIsing}) and (\ref{dif1}), it is easy to obtain for the non vanishing elements of the $4\times4$ matrix $U_p(t)$ the differential equation
\begin{eqnarray}
i\frac{d}{dt}\left(
\begin{array}{cc}
U_p^{11}(t)&U_p^{12}(t)\\
U_p^{21}(t)&U_p^{22}(t)
\end{array}
\right)=&\nonumber\\
 \left(
\begin{array}{cc}
U_p^{11}(t)&U_p^{12}(t)\\
U_p^{21}(t)&U_p^{22}(t)
\end{array}\right)\left(
\begin{array}{cc}
h(t)&-i\sin\phi_p\\
i\sin\phi_p&2\cos\phi_p-h(t)
\end{array}\right)&\; 
\end{eqnarray}
and $U_p^{33}(t)=U_p^{44}(t)=e^{-it\cos\phi_p}$.
From the two-coupled first order differential equations, it is easy to obtain the decoupled second order ones. For example, one has for the $11$-component
\begin{equation}
i u''-2\cos\phi u'-(h'-i\sin^2\phi+ih[2\cos\phi-h])u=0
\label{dif11} 
\end{equation}
with $u(0)=1$, $u'(0)=-ih(0)$ and similar equations for the other components. Taking for the field the form $h(t)=h_1+h_2\cos(\omega t)$ and putting it in the differential equation (\ref{dif11}) one gets solutions in terms of double confluent Heun function  $\rm H_D$, namely for $U_p^{11}(t)$ and $U_p^{12}$ one has
\begin{eqnarray}
U_p^{11}(t)=\frac{1}{2h_2}\left[(\cos\phi_p-h_1)e^{iA_t}{\rm H_D}(\alpha,\beta,\gamma,\delta,-iz)\right. \nonumber\\
\left. -(\cos\phi_p-2h_2-h_1)e^{-iA_t}{\rm H_D}(-\alpha,\beta,\gamma,\delta,iz)\right]\; ,
\end{eqnarray} 
\begin{eqnarray}
U_p^{12}(t)=\frac{i\sin\phi_p}{2h_2} \left[e^{iA_t}{\rm H_D}(\alpha,\beta,-\gamma,\delta,-iz)\right. \nonumber\\
\left. -e^{-iA_t}{\rm H_D}(-\alpha,\beta,-\gamma,\delta,iz)\right]\; ,
\end{eqnarray} 
with $z=\tan (\omega t/2)$, $A_t=\frac{h_2}{\omega}\sin(\omega t)-t\cos\phi_p$, $\alpha=\gamma/2=4h_2/\omega$, 
$\beta=\frac{4}{\omega^2}\left[h_2^2-|h_2-h_1+e^{i\phi_p}|^2  \right]$ and 
$\delta=-\frac{4}{\omega^2}\left[h_2^2-|h_2+h_1+e^{i\phi_p}|^2  \right]$, and similar expressions for the other components.

\begin{figure} [ht]
        \includegraphics[width=8cm]{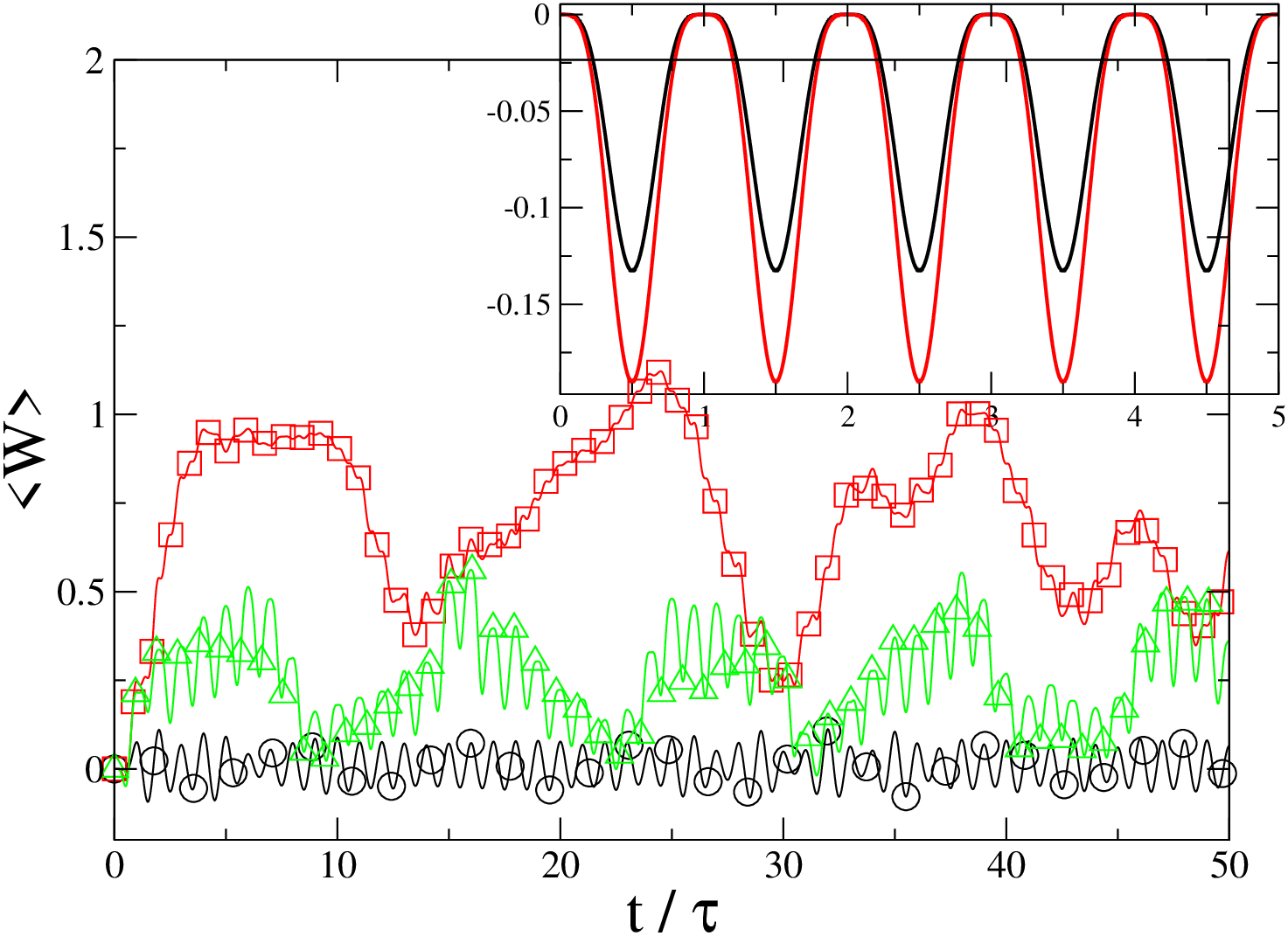}
       \caption{Time evolution of the average work for the Ising chain with a sine square periodic transverse field at $\beta=1$ and maximal amplitude $h_o=1/2$ with periods $\tau=1$ (circles), $\tau=3$ (squares) and $\tau=4$ (triangles). The inset shows the free energy difference (lower curve) and the average work (top curve) for a large period $\tau=50$ of the field. In this case the transformation is nearly adiabatic.    
        \label{Fig8} }
\end{figure}

The equilibrium initial state is factorized as $\rho(0)\propto\prod_{p=1}^{N/2}e^{-\beta \tilde{H}_p(0)}$ and its time-evolution is given by the tensor product
\begin{equation}
\rho(t)=\rho_1(t)\rho_2(t)...\rho_{N/2}(t)
\end{equation}
with 
\begin{equation}
\rho_p(t)=U_p(t)\rho_p(0)U^+_p(t)\; .
\end{equation}
The average work performed on the system during the time $t$ with the periodic forcing is given by the integral of the magnetization:
\begin{equation}
\langle W\rangle(t) =-\int_0^t\! {\rm d}s\; h'(s) Tr\left\{ \rho(s) M^z \right\}
\end{equation}
where $M^z=\sum_{p=1}^{N/2} M_p^z$, with $M_p^z=a^+_pa_p+a^+_{-p}a_{-p}-1$, is the total magnetization operator in the $z$-direction. Using the factorised form of $\rho(t)$ and the additive structure of $M^z$, it is easy to compute $\langle M^z\rangle(t)$ as a sum of $N/2$ independent modes. One obtains
\begin{eqnarray}
\langle M^z\rangle(t)&=&\sum_{p=1}^{N/2}
\frac{\tanh\left(\frac{\beta\Lambda_p}{2}\right)}{\Lambda_p}
\big[(h(0)-\cos\phi_p)(2|U_p^{11}|^2-1)\nonumber \\
& -& 2\sin\phi_p Im\left\{U_p^{12}{U_p^{11}}^* \right\}\big]
\end{eqnarray}
with $\Lambda_p=\sqrt{\sin^2\phi_p+(h(0)-\cos\phi_p)^2}$.
Together with the solutions of the $U_p$ components in terms of Heun functions, this formally solves the problem for the average work. One may notice that the factor $\tanh(\frac{\beta \Lambda(0)}{2})$ in this expression leads to the previously observed $\tanh(\beta/2)$ in (\ref{Wtanh1}) since in that case $h(0)=0$ and then $\Lambda_p(0)=1$ $\forall p$ and one has for the average magnetization
\begin{eqnarray}
\langle M^z\rangle(t)=-\tanh\left(\frac{\beta}{2}\right)[1+2\sum_p\cos\phi_p |U_p^{11}|^2\nonumber \\
+ \sin\phi_p Im\left\{U_p^{12}{U_p^{11}}^*\right\}]\; 
\end{eqnarray}
and consequently the average work can be written as in (\ref{Wtanh1}): $\langle W\rangle_\beta= \tanh(\beta/2)\langle \delta W\rangle$. 
The independence of the $N/2$ modes also explains the linear behavior of the variance 
observed in the previous section since $\langle W^2\rangle_\beta-\langle W\rangle_\beta^2=\sum_q^{N/2}\langle W_q^2\rangle_\beta-\langle W_q\rangle_\beta^2 \sim  N$, where $W_q$ is the work associated to the $q$th mode. 
Finally, taking $h_1=-h_2=h_o/2$, such that $h(t)=h_o\sin^2(\frac{\omega t}{2})$, we have an exact solution for a situation which closely resembles to the triangular drift treated above numerically. We show the behavior of the average work on figure \ref{Fig8} for different periods $\tau=2\pi/\omega$. As previously observed, the average work fluctuations are maximum around $\tau=3$.

\subsubsection{XX chain in a longitudinal magnetic field}
\begin{figure} [ht]
        \includegraphics[width=8cm]{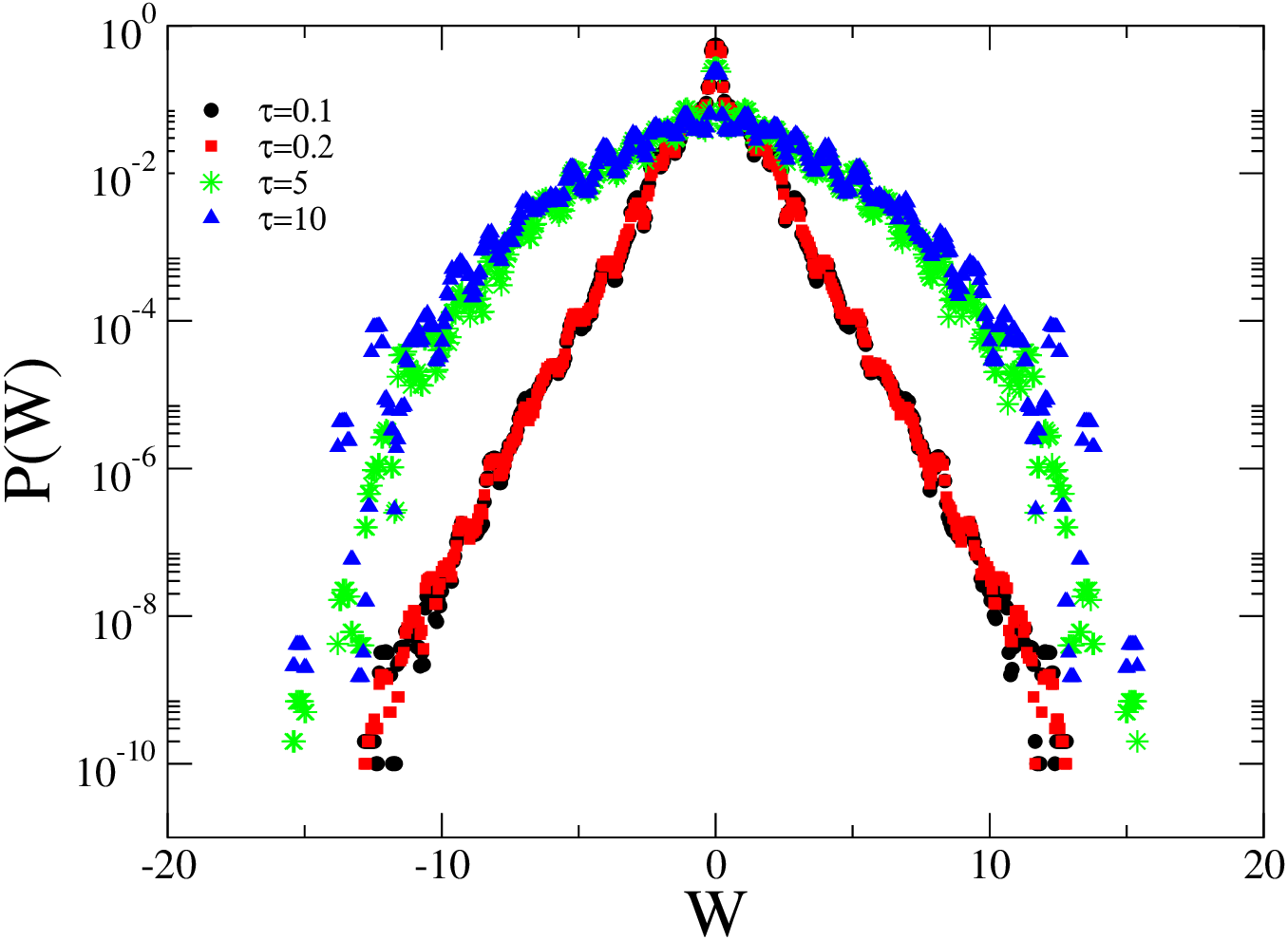}
         \caption{Work probability distribution  for the XX quantum chain with $N=8$ spins at $\beta=0$ for $h_o=1/2$.     \label{Fig9}  }
\end{figure}

\begin{figure} [ht]
        \includegraphics[width=8cm]{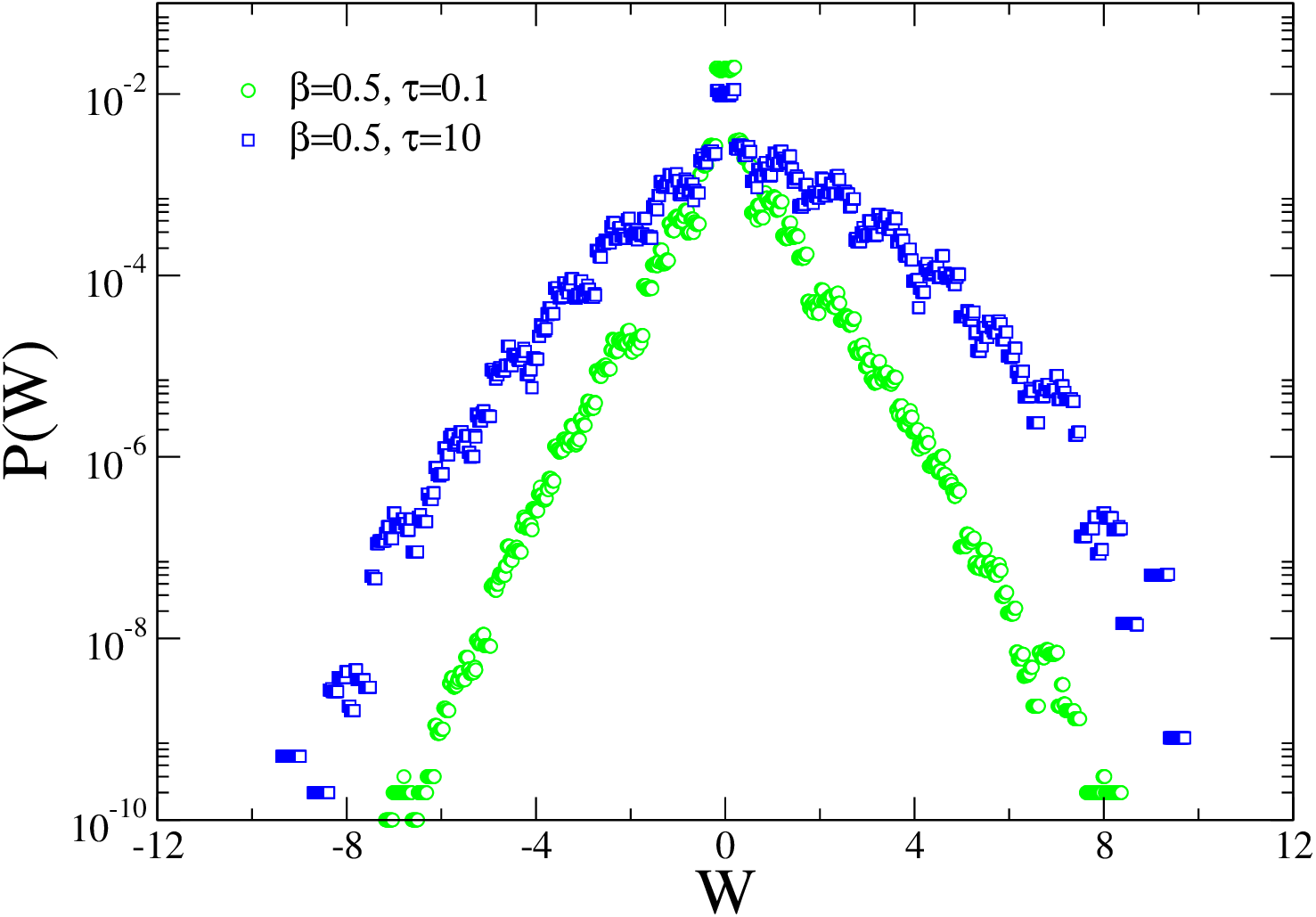}
         \caption{Work probability distribution  for the XX quantum chain with $N=8$ spins at  $\beta=1/2$ for $h_o=1/2$.     \label{Fig9bis}  }
\end{figure}

Finally we present the results obtained in the forced regime for the $XX$ chain. In figure \ref{Fig9} we show the stationary work probability distributions for short and long period $\tau$ on a chain of $N=8$ spins obtained from an initial infinite temperature state after applying  the field over typically several tens of periods (after about $30$ periods the work distributions are collapsing nicely on the same curve). We see clearly the appearance of two different regimes depending on the time scale $\tau$. For very short periods $\tau\ll 1$, the shape of the distribution is very well approximated by an exponential law $e^{-|W|/\alpha}/(2\alpha)$ while for long enough time scale $\tau$ its shape is close to a normal law. One may notice on figure \ref{Fig9} the very good collapse of the distributions for different values of the period $\tau$ in both short and long periods regimes.
We have also seen that this transition survives at finite temperature. The main difference between the infinite and the finite temperature cases being that in the last case the work  distribution is no more symmetric but squeezed toward the positive work values as seen in figure \ref{Fig9bis}. 
 
\begin{figure} [htb]
\vskip0.9cm
      \includegraphics[width=8cm]{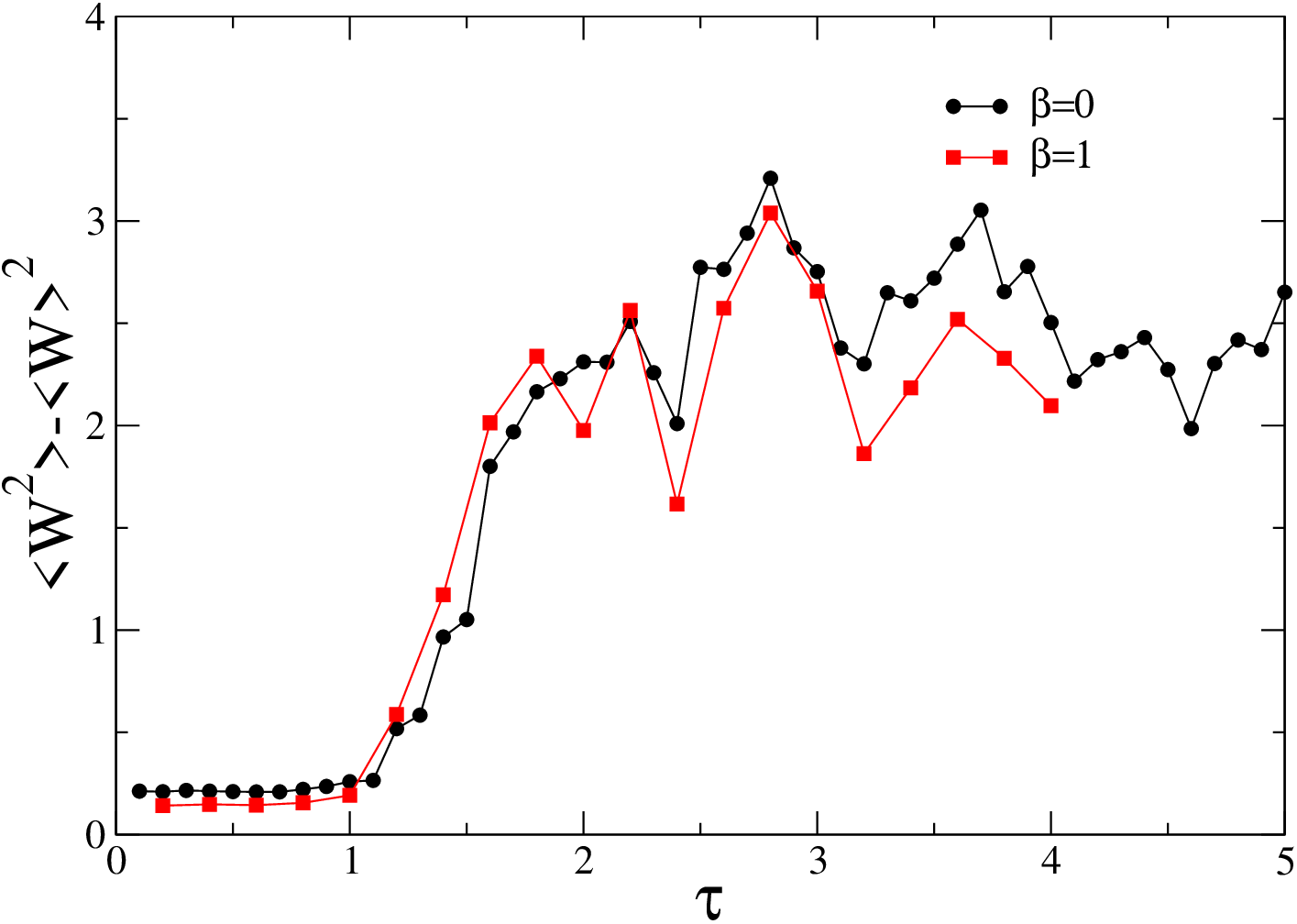}
        \caption{Variance of the work distribution for the XX quantum chain with $N=8$ spins at $\beta=0$ and $\beta=1$ for $h_o=1/2$ as a function of the field period $\tau$.
        \label{Fig10}  }
\end{figure}

\begin{figure} [htb]
\vskip0.7cm
      \includegraphics[width=8cm]{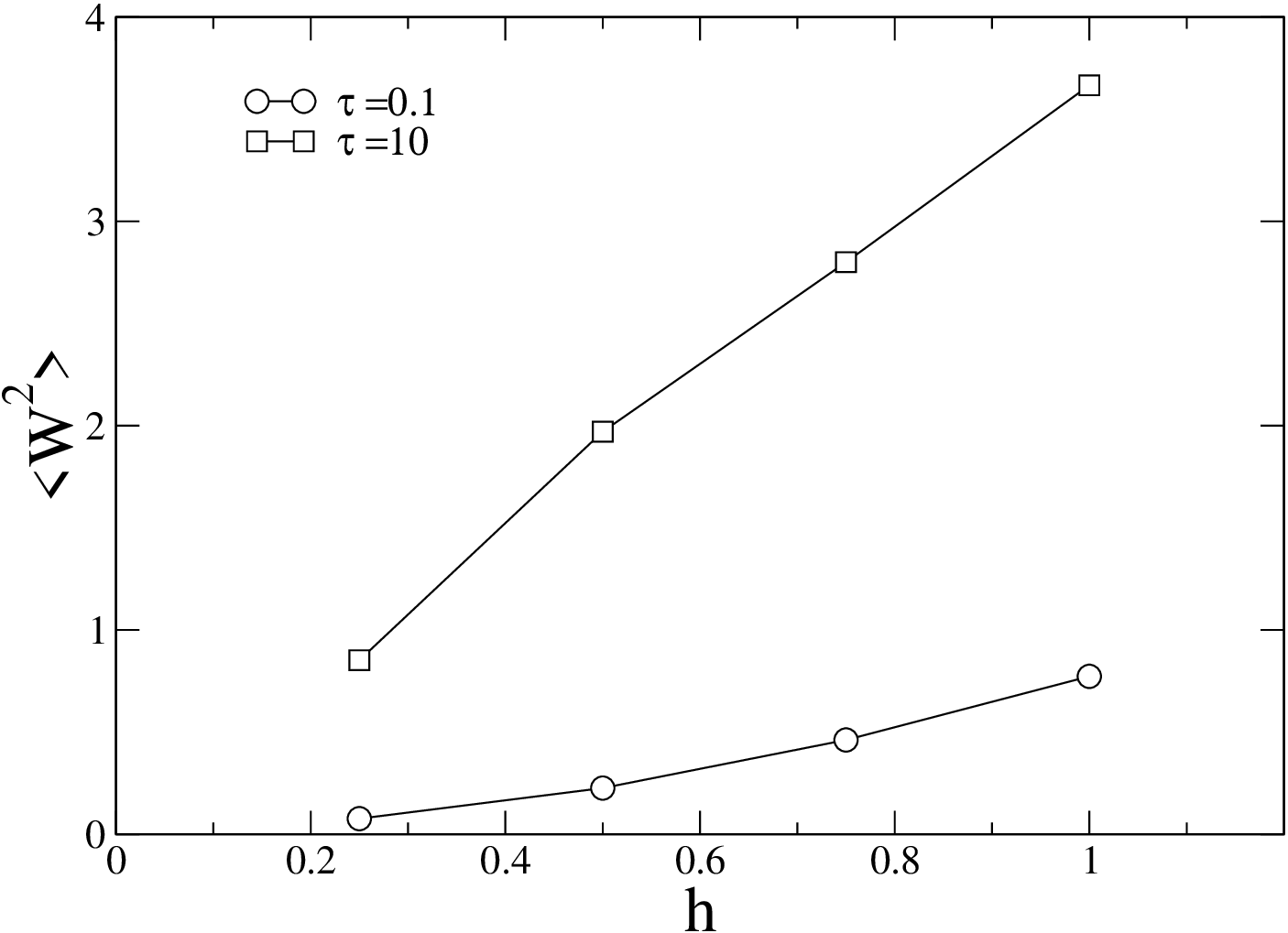}
        \caption{Second moment of the work distribution for the XX quantum chain with $N=8$ spins at $\beta=0$ as a function of the field maximal amplitude $h_o$.
        \label{Fig11}  }
\end{figure}

\begin{figure} [ht]
\vskip0.9cm
      \includegraphics[width=8cm]{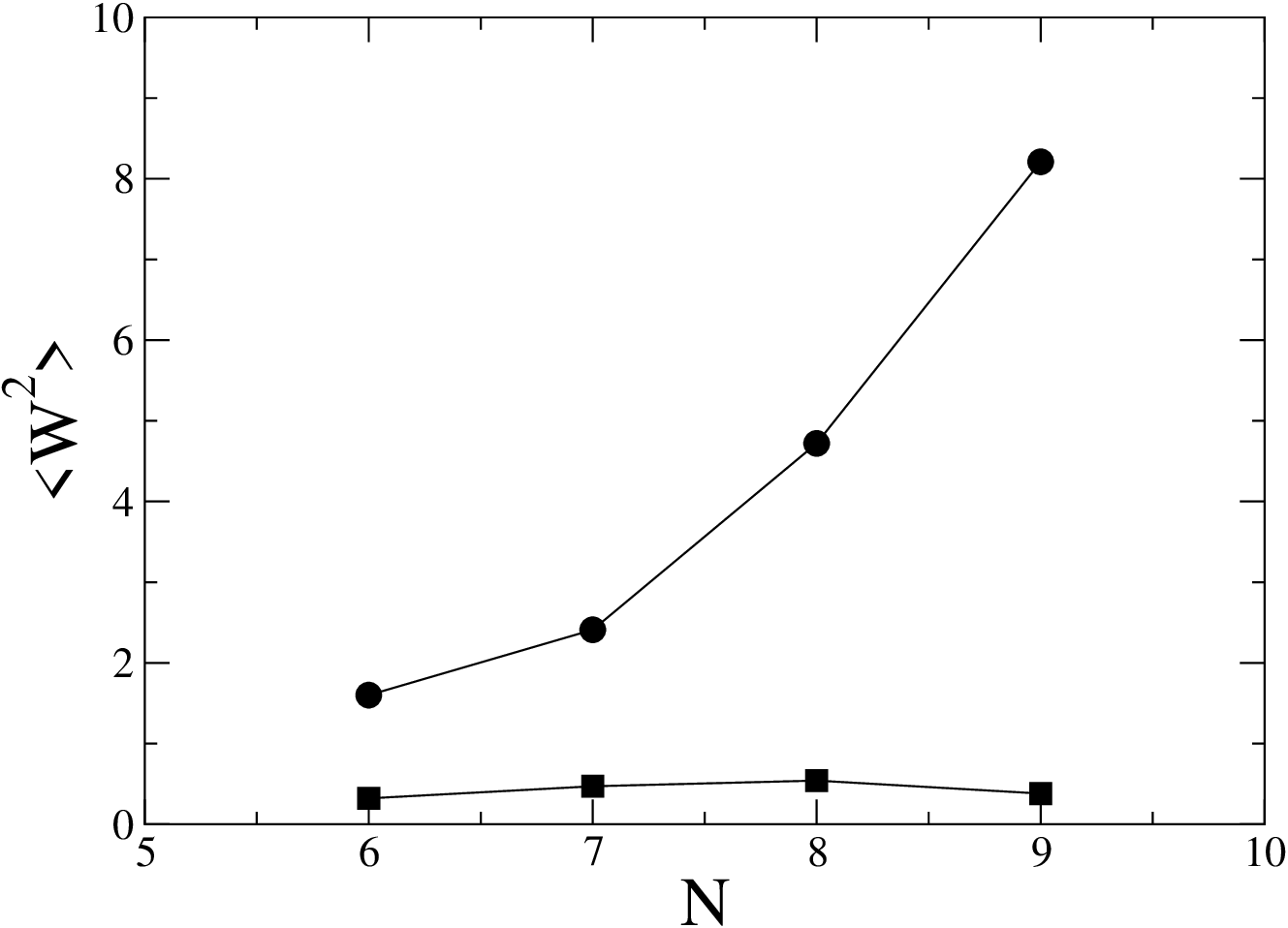}
        \caption{Second moment of the work distribution for the XX quantum chain with $h_o=1/2$, at $\beta=0$ as a function of the system size $N$ for $\tau=0.1$ (squares) and $\tau=10$ (circles).
        \label{Fig12}  }
\end{figure}

To give evidences of this transition, we have plotted on figure \ref{Fig10} the variance $\langle W^2\rangle-\langle W\rangle^2$ (the average work is vanishing for $\beta=0$) of the distribution $P(W)$ as a function of the time scale $\tau$. For a system of size $N=8$, we observe a sharp increase of the variance from values around $0.25$ at time scales smaller than for $\tau_c\simeq 1$ to values around $2.5$ at larger periods $\tau$ where the shape is gaussian-like. Note that the threshold value $\tau_c$ is depending on the system size, decreasing as the size is increased. 
On figure \ref{Fig11} we show the field amplitude dependence of the variance $\langle W^2\rangle_0$ of the work distribution for short and long periods of the external drift. One may notice that at long periods, the increase of the variance with the field amplitude is almost linear, while it is more curved at short periods.   
On figure $\ref{Fig12}$ we see that as the size is increased the variance at large time scales strongly grows with N, while there is almost no dependence on the size at high frequencies. It means that when the period of the external field is to short, the system has no time to follow the perturbation and only transitions between nearby levels are significantly induced. For the density of work $w=W/N$ this leads to a $\delta(w)$ distribution in the limit $h\rightarrow 0$, since as seen on figure~\ref{Fig11}, $\alpha=\sqrt{\langle W^2\rangle/2}$ is a vanishing function of the amplitude of the field $h$, presumably linear. On the contrary, and as the system size is increased, at long periods enough, meaning that the external field drives efficiently the system from its initial level to a new state, transitions between far apart levels are becoming significant. Taking that the variance seems to be $const. h N^2$ from figures~\ref{Fig11} and \ref{Fig12},  the distribution $\pi(w)$ in that case behaves as $e^{-const. w^2/h}$. 

\begin{figure} [ht]
\vskip0.8cm
      \includegraphics[width=8cm]{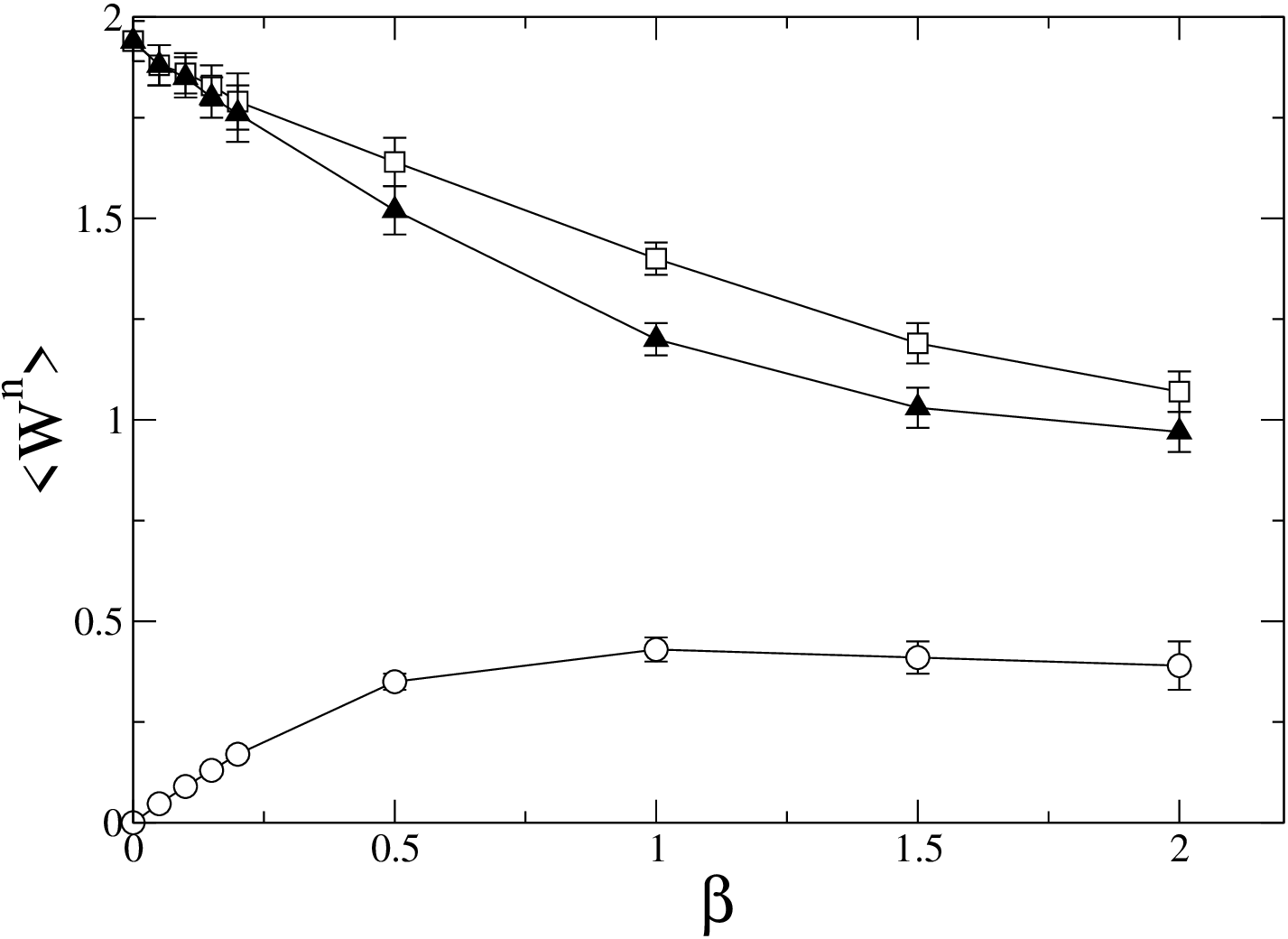}
        \caption{First (circles), second (squares) and variance (triangles) of the work distribution for the XX quantum chain with $N=8$ spins at $\tau=10$ at $h_o=1/2$ as a function of the inverse temperature $\beta$.
        \label{Fig13}  }
\end{figure}

Finally, we have analyzed the temperature dependence of the work distribution $P(W)$
in the large $\tau$ regime. As seen on figure \ref{Fig13}, the average work after a first increase with $\beta$ seems to finally saturate at a value close to $0.44$. We have checked that this shape is compatible with a $\tanh(\mu\beta)$ behavior. Contrary to the almost temperature independent variance of the Ising case distributions, we see here that the variance of the distribution is decreasing(increasing) with $\beta$(temperature), which is what one would normally expect.

\section{Summary and discussion}
In this study we have presented the numerical and analytical results we have obtained for the work distribution of small quantum systems driven by an external field. We have considered two different models, namely the integrable Ising quantum chain in a time-dependent transverse field and the $XX$ quantum chain in a time-dependent longitudinal magnetic field.   In this last model, the presence of the longitudinal field breaks the free fermionic structure of the chain, while the transverse field in the Ising case preserves the free particles structure which obviously leads to integrability. 
In a first stage, we have checked the validity of both quantum Jarzynski and quantum Crooks relations that were largely discussed in section 2. 
After this initial check, validating our numerical approach, we focused our attention to the periodic driven situation. In the Ising chain, exploiting its free fermionic structure, we solved exactly the unitary dynamics of the system in terms of somehow complicated double confluent Heun functions. The main features observed are, as expected, that for a slow enough process, that is large period of the oscillations of the field compared to the coupling constant, we recover the adiabatic situation where the work slightly differs, from above, from the equilibrium free energy difference. This is a statement of the second principle. As the process is fastened, the average work starts to deviate significantly from the free energy difference with fluctuations in time that are growing as the frequency of the field is increased. Nevertheless, at very high frequencies, the work fluctuations decrease again to zero. In this case, the variation of the field is so fast that the system is not able to follow it anymore. The maximum amplitude of the work fluctuations is obtained at an intermediate period which is of order of the inverse field amplitude, at least for large field enough. 
This threshold is understood as a dynamical resonance. In the Ising chain, one is not able to observe a steady work distribution. 
The behavior of the $XX$ chain in a longitudinal periodic field is quite different. We observe numerically, that the long-time work distribution reaches a steady shape which is strongly dependent on the frequency of the field. Indeed, at high frequencies the distribution is well fitted by an exponential curve, $e^{-|W|/\alpha}/(2\alpha)$, where $\alpha\sim h$ is almost size-independent. Lowering the frequency, there is a sharp transition toward a Gaussian like behavior for the work distribution with a variance which seems from the numerics to be proportional to $h N^2$, . 
In order to understand the transition between these two limit distribution, one has to realize that the energy band of the chain has a width $\delta_E$ of order $N$ (since the typical coupling $J$ is set to one). The periodic perturbation introduces a typical energy scale $\omega\equiv 2\pi/\tau$. For $\omega> \delta_E$, there is no resonant coupling of the system with the periodic forcing. But as soon as $\omega< \delta_E$ resonant coupling leads to the appearance of resonant peaks, that is sharp increases of the transition probability in the regions $W=\pm \omega$ and integer multiples of $\omega$. The typical width of those peaks is of the order of the amplitude $h$ of the perturbation. Consequently, the deviation to the initial exponential distribution starts at periods $\tau > {\cal O}(2\pi/N)$. As the period is increased further, the first peaks moves toward the center of the distribution and new resonant peaks enter into play from the boundaries. Finally, the superposition of these resonant peaks for $\tau>{\cal O}(1/h)$ leads to a new limit distribution of the work. So the transition region between the two limit distributions is ${\cal O}(2\pi/N)< \tau < {\cal O}(1/h)$. 
Finally, one may remark that the change of shape from exponential-like to Gaussian-like could be linked to the integrability of the model. Indeed, at very short periods, the system is not able to follow the external perturbation and its behavior is governed by the initial integrable Hamiltonian ($XX$-chain without field) while at larger periods the system feels effectively its non-integrability, possibly leading to a Gaussian distribution of level spacing (which is our work). To confirm eventually this scenario one needs to push further this investigation.

\bibliography{FluctPR5}

\end{document}